# Two-dimensional simulation of Argon dielectric barrier discharge (DBD) in plasma actuator structure with COMSOL Multiphysics

Ramin Mehrabifard

Department of Physics and Institute for Plasma Research, Kharazmi University, 49 Dr. Mofatteh Avenue, Tehran, Iran

## Abstract

Dielectric barrier discharge (DBD) plasma is used for various applications. DBD is also one of the most efficient and low-cost methods for active fluid flow control. In this study, a detailed physical model of DBD in atmospheric pressure at 1kV DC voltage is developed with COMSOL Multiphysics software. Argon gas is also used as a background gas and electrodes are assumed to be copper. Plasma parameters such as electron and ion density, electric field, potential, and temperature for different distances of electrodes (1.0mm, 0.9mm, 0.8mm) have been investigated. Moreover, the effect of dielectric type (Quartz, Silica Glass, Mica) on these key parameters is investigated. The results of the simulation show that the longitudinal distance of the buried electrodes to the exposed electrodes has a direct influence on parameters such as electron temperature, and electron and ion density which are the main factors of fluid flow control. These parameters have the maximum value when mica is used as a dielectric and the lowest value when silica glass is utilized.

**Keywords**: Dielectric Barrier Discharge, Argon Plasma, Plasma Actuator, Body Force, COMSOL Multiphysics, Gas Discharge

## 1. Introduction

Due to its exceptional features, cold atmospheric pressure plasma (CAP) has been applied in a wide range of scientific fields (Da Ponte et al., 2012; Hati et al., 2012; Lukes et al., 2014; Mehrabifard et al., 2020; Zimmermann K.; Shimizu, Tetsuji; Morfill, Gregor E.; Wolf, A.; Boxhammer, Veronika; Schlegel, Jürgen; Gansbacher, Bernd; Anton, Martina, 2011). CAPs can involve partially ionized gas, many active species (positive/negative ions, radicals) UV radiation and, transient electric field (Graves, 2012; Mehrabifard et al., 2017; Weltmann et al., 2010). Controlling the separation of flow with the use of atmospheric cold plasma is considered very promising (Jayaraman, Lian, et al., 2007; Shang & Huang, 2014; Sohbatzadeh et al., 2018). The separation of the flow from the surface will occur at high angles of attack in a variety of cases, including diffusers, cars, feathers of turbines and airfoils, and will always be accompanied by losses, which will reduce significantly the performance parameters and efficiency. The cold plasma can control this flow by making the body force produced by the difference in charge particle and the electric field in the system (Boeuf et al., 2007). Several methods are used to obtain the body force created by the non-equilibrium plasma (Abdollahzadeh et al., 2012). Different models of atmospheric pressure discharge are also used to create body force including direct discharge, microwave discharge, corona discharge, spark and dielectric barrier discharge (Boeuf et al., 2007; Georghiou et al., 2005; Shang & Huang, 2010). Therefore, the dielectric barrier discharge is one of the most common methods for creating body force (Jayaraman, Cho, et al., 2007). In the plasma actuator, the structure of the electrodes is non-symmetrical that are located two sides of the dielectric. Many researchers have used numerical codes to simulate plasma. Studies have shown that in previous work, due to the use of numerical codes, the number of reactions considered for discharge and surface reactions are limited (Soloviev & Krivtsov, 2009).

In this research, all the characteristics of an electric discharge of argon gas in a two-dimensional structure used in a plasma actuator have been evaluated. And the effect of the distances of the electrodes from each other and the dielectric material in which the grounded electrode is buried on it, on the body force main parameters (net charge and electric field magnitude) has been investigated.

## 2. Plasma model

### 2.1. Geometry and Description

The general schematic of the actuator is shown in Fig. 1. Copper electrodes dimensions are ($0.2mm \times 3mm$) and the dielectric dimensions are ($0.6mm \times 14mm$). The gas used in this structure is argon, which contains neutral particles, electrons, negative ions, and positive ions. The total space around the structure involves argon. The input voltage leads to the production or consumption of each component. Argon reacts in different forms on the surface, which is among the seven probable reactions that are shown in Table 1.

In most of the studies that have been done in this field, pure gases have been used to produce smaller components and simplify computations. Increasing reactions also result in longer computing time. In addition to the seven reactions mentioned above, two surface reactions are considered, as shown in Table 2.

Table1.Table of collisions and reactions modeled.

| Reactions | Formula | Type | $\Delta\varepsilon(eV)$ |
|---|---|---|---|
| *1* | $e+Ar \rightarrow e+Ar$ | *Elastic* | *0* |
| *2* | $e+Ar \rightarrow e+Ars$ | *Excitation* | *11.5* |
| *3* | $e+Ars \rightarrow e+Ar$ | *Superelastic* | *-11.5* |
| *4* | $e+Ar \rightarrow 2e+Ar^{+}$ | *Ionization* | *15.8* |
| *5* | $e+Ars \rightarrow 2e+Ar^{+}$ | *Ionization* | *4.24* |
| *6* | $Ars+Ars \rightarrow e+Ar+Ar^{+}$ | *Penning ionization* | - |
| *7* | $Ars+Ar \rightarrow Ar+Ar$ | *Metastable quenching* | - |

Table2. Table of surface reactions

| Reaction | Formula | Sticking Coefficient |
|---|---|---|
| 1 | $Ars \rightarrow Ar$ | *1* |
| 2 | $Ar^{+} \rightarrow Ar$ | *1* |

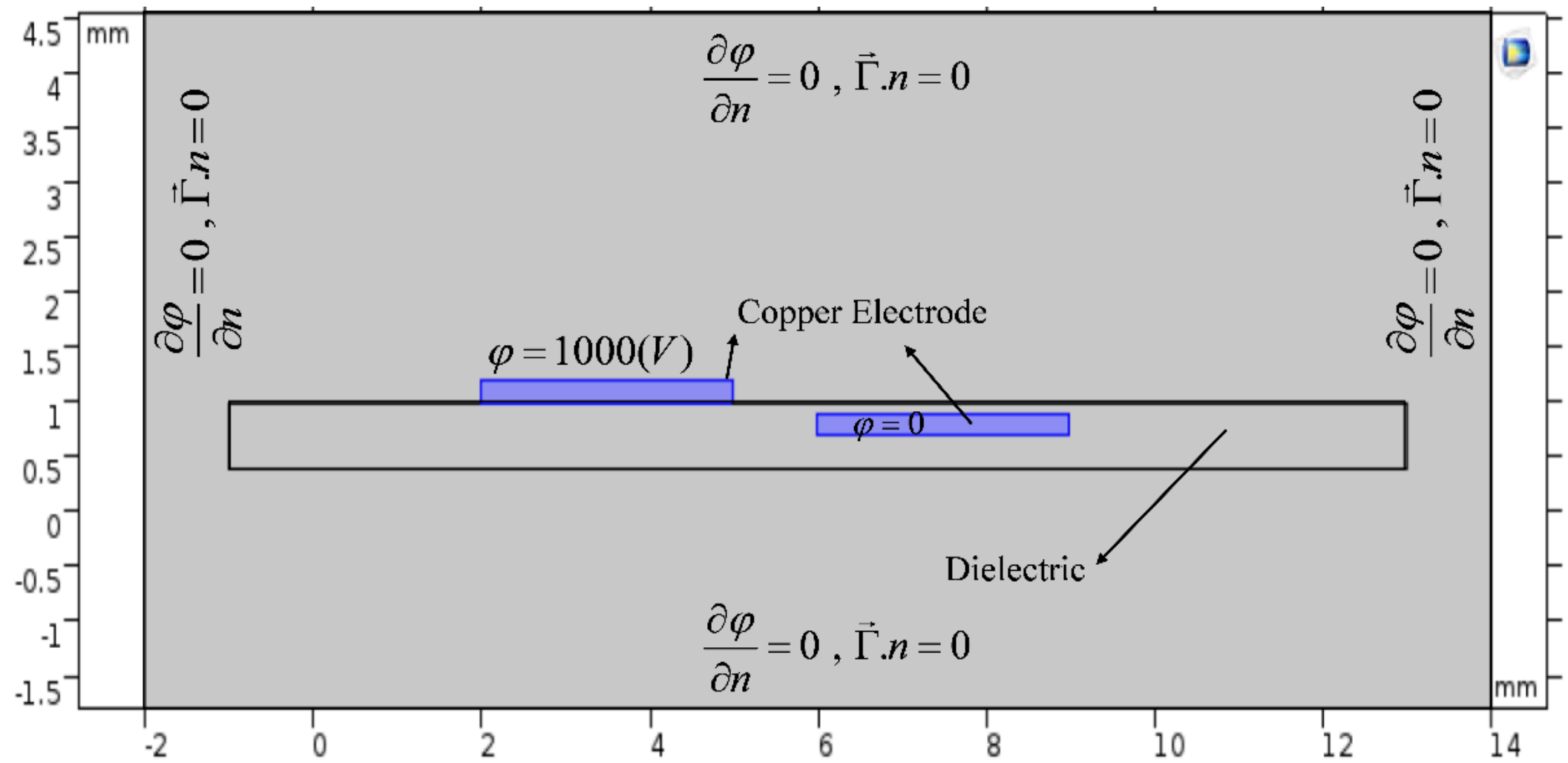

Fig 1. Plasma actuator structure and boundary condition for potential

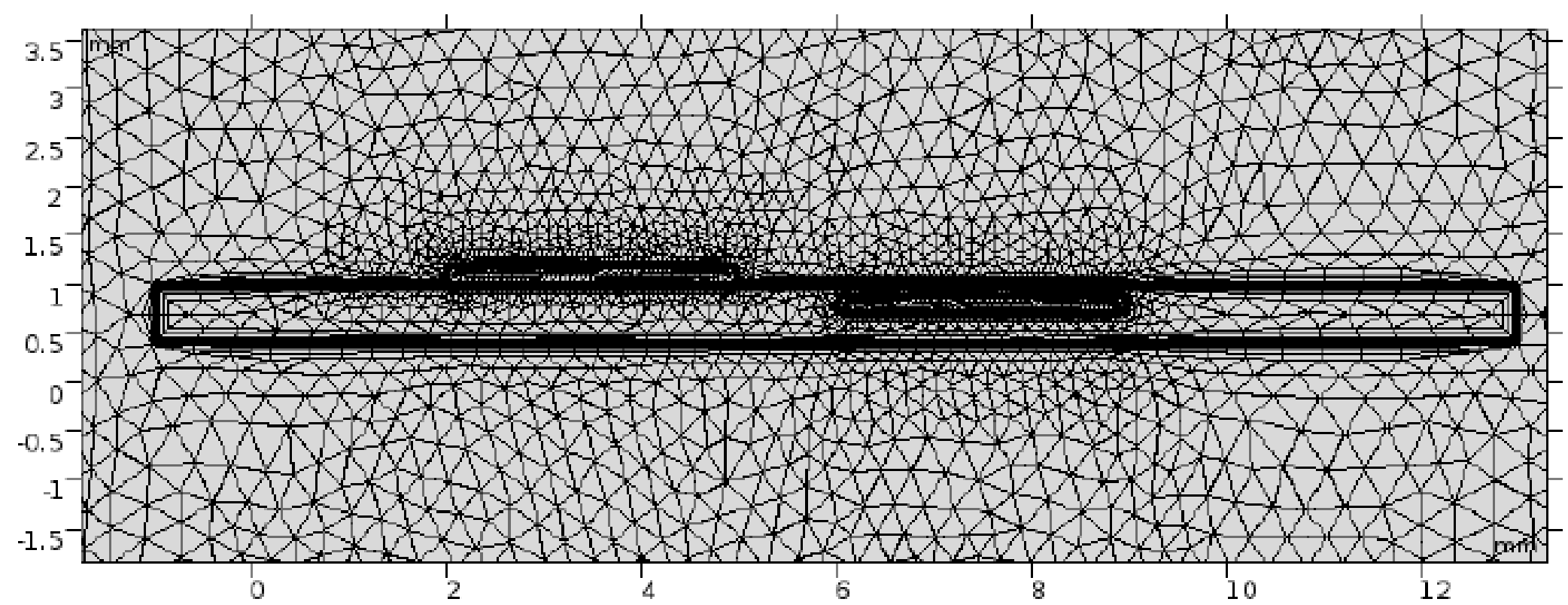

Fig 2. Mesh and its density at the sharp edge

The fluid model is used in this study. By solving the drift-diffusion equations, the density of electrons and energy will be obtained. The governing equations for electric discharge with the drift-diffusion approximation are as follows:

$$\frac{\partial n_e}{\partial t} + \nabla.\vec{\Gamma}_e = R_e - (\vec{u}.\nabla)n_e \qquad (1)$$

$$\vec{\Gamma}_e = -(\vec{\mu}_e.\vec{E})n_e - \vec{D}_e.\nabla n_e \qquad (2)$$

Equation (1) expresses the electron continuity equation. $n_e$ is the electron density, $D_e$ is electron diffusion coefficient, and $\Gamma_e$ electron flux, and $u$ is average velocity of the species and $R_e$ is the electron production rate. Equation (2) is the equation of electron flux, which consists of two parts: drift and diffusion. The electron energy density is calculated by the following equation:

$$\frac{\partial n_\varepsilon}{\partial t} + \nabla.\vec{\Gamma}_\varepsilon + \vec{E}.\vec{\Gamma}_\varepsilon = R_\varepsilon - (\vec{u}.\nabla)n_\varepsilon$$
$$\vec{\Gamma}_\varepsilon = -(\vec{\mu}_\varepsilon.\vec{E}).n_\varepsilon - \vec{D}_\varepsilon.\nabla n_\varepsilon \qquad (3)$$

This expression $\vec{E}.\vec{\Gamma}_\varepsilon$ represents the amount of energy obtained form an electron by electric field. $R_e$ is the energy derived from non-elastic collisions calculated by the following equation:

$$R_\varepsilon = S_{en} + \frac{Q + Q_{gen}}{q} \qquad (4)$$

$S_e$ is power dissipation, $Q_{gen}$ is the thermal source and $q$ is electron charge. $D_e$ is electron diffusion coefficient, $\mu_\varepsilon$ energy mobility, and $D_\varepsilon$ energy distribution coefficient.

$$D_\varepsilon = \mu_\varepsilon T \quad D_e = \mu_e T_e \quad \mu_\varepsilon = \frac{5}{3}\mu_e \qquad (5)$$

We used the Townsend coefficients of the electron source that is calculated by the following equation:

$$R_e = \sum_{j=1}^{M} x_j a_j N_n \left|\Gamma_e\right| \qquad (6)$$

$M$ is number of reactions, $x_j$ the molar fraction of the target species for the $j$ reaction, $a_j$ the Townsend coefficient for the reaction j, and $N_n$ is the total number of neutral particles. Considering the number $p$ of non-elastic electron collisions, we will have:

$$R_\varepsilon = \sum_{j=1}^{p} x_j a_j N_n \left|\Gamma_e\right| \Delta\varepsilon_j \qquad (7)$$

Which $\Delta\varepsilon_j$ is the energy dissipation of j reaction. In non-electron species, the above equations are used for mass fraction:

$$\rho\frac{\partial w_k}{\partial t} + \rho(\vec{u}.\nabla)w_k = \nabla.\vec{j}_k + R_k \qquad (8)$$

$w_k$ is the ionic density, $j_k$ is the energy flux of the ions. The electrostatic field is obtained by the following equation:

$$\nabla.(\varepsilon_0\varepsilon_r E) = \rho \qquad (9)$$

$\varepsilon_0$ is the permittivity of vacuum, and $\varepsilon_r$ is a relative dielectric constant.

### 2.2. Boundary condition

With respect to the boundary conditions for the electron flux and energy flux, the following relation is established:

$$-\hat{n}.\vec{\Gamma}_e = (\frac{1}{2}\upsilon_{eth}n_e) - \sum_p \gamma_p(\vec{\Gamma}_p.\hat{n}) \tag{11}$$

$$-\hat{n}.\vec{\Gamma}_\varepsilon = (\frac{5}{6}\upsilon_{eth}n_e) - \sum_p \varepsilon_p\gamma_p(\vec{\Gamma}_p.\hat{n}) \tag{12}$$

The right-hand side of equation (11) shows the electron production by the secondary electron and $\gamma$ is the secondary electron coefficient. On the surface of electrodes, ions and excited species are neutralized by the surface reaction. Surface interactions on the electrode are indicated by the $\beta_j$ coefficient, which indicates the probability of the function of the j species. The continuity equation for ions is defined as follows:

$$\frac{\partial n_i}{\partial t} + \nabla.(n_i\vec{u}) = -\nabla.(\mu_i n_i q_i \nabla\varphi - D_i\nabla n_i) + S_i \tag{13}$$

$\varphi$ And $S_i$ indicate the electrostatic potential and the electron density change rate.

## 3. Result and discussion

The evolution of plasma characteristics by buried electrode displacement are investigated in this study. The transverse distance to the dielectric edge is 0.2 mm and the longitudinal distance from the electrode are considered 0.8*mm*, 0.9*mm* and 1 *mm*. The boundary conditions for free space are Neumann and, on the electrode, surface are Dirichlet conditions (fig 1). Figure 3 shows the electrical potential in the presence of plasma at different distances of the electrodes. The power electrode voltage is 1000 V (DC) and the buried electrode is at zero potential (ground).

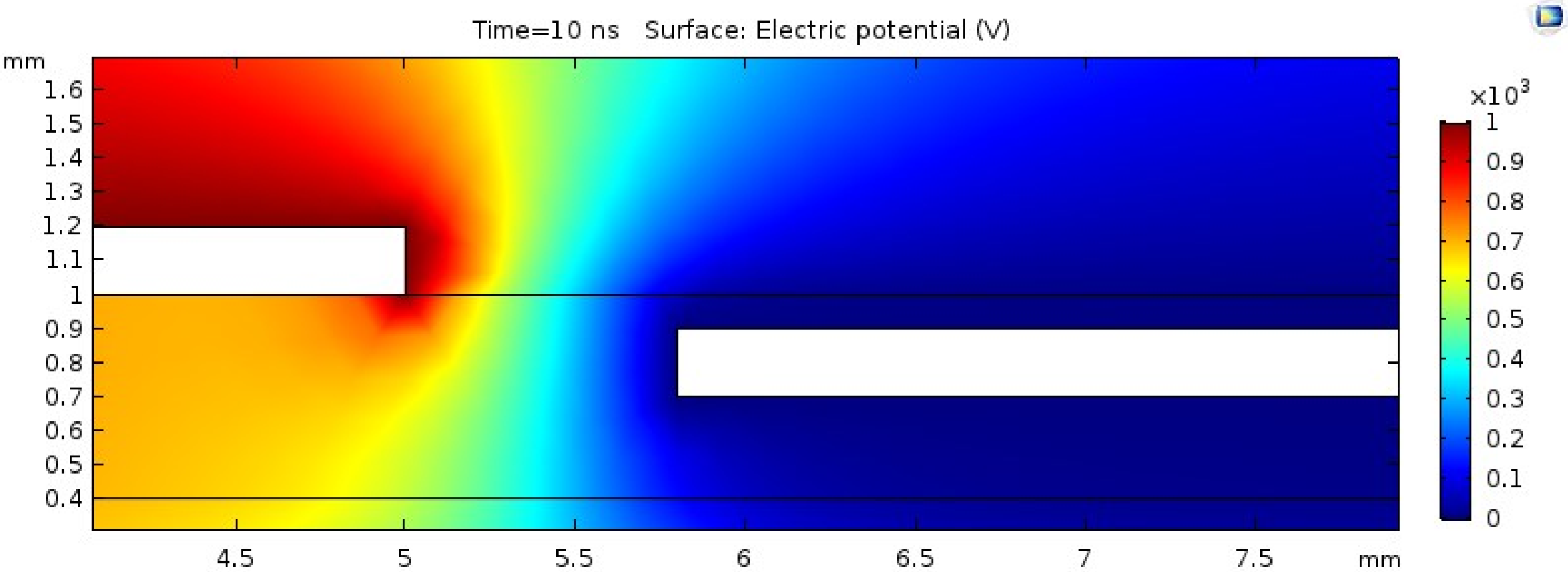

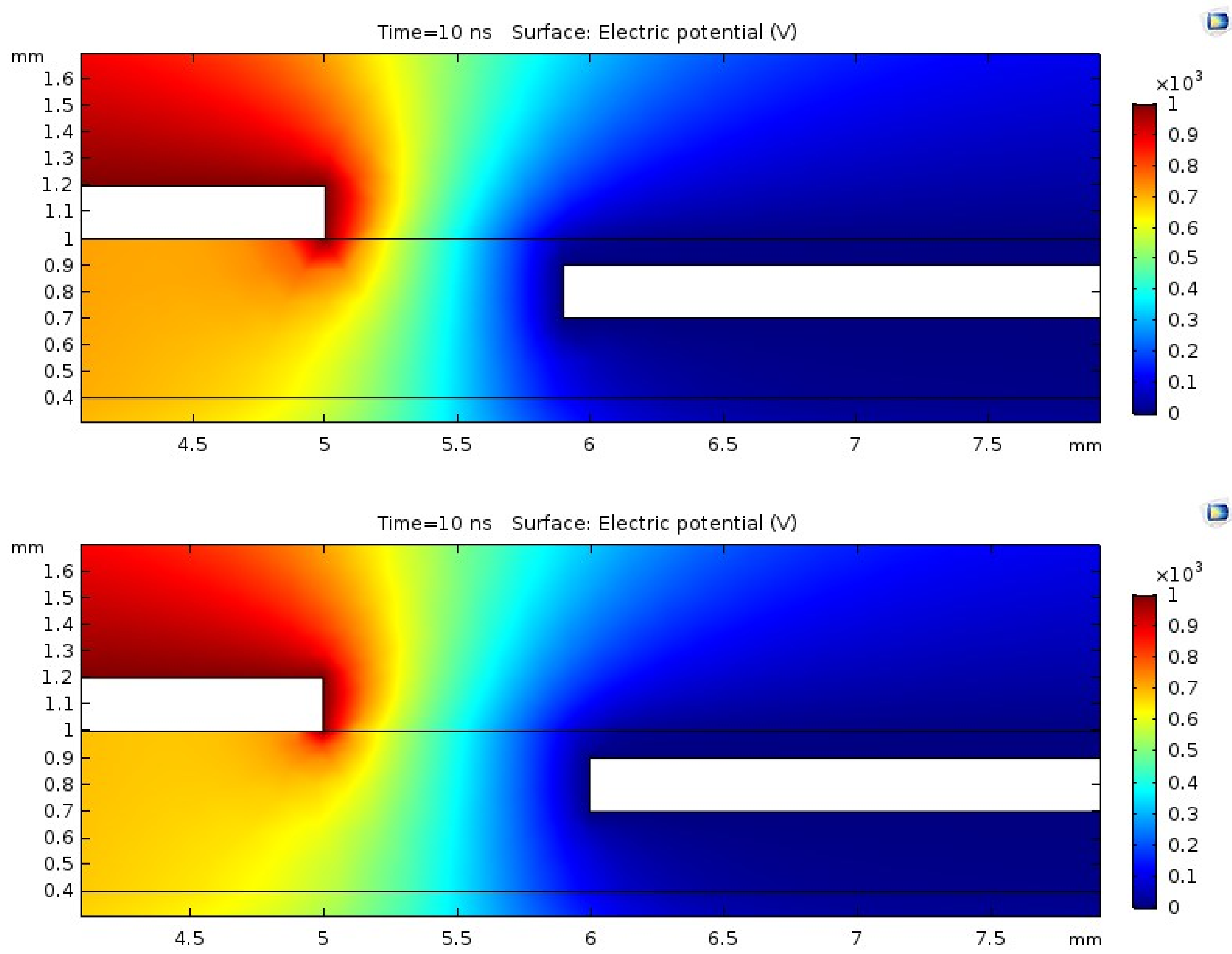

Fig 3. Electric potential distribution on the cross-section of the actuator for different distances of electrodes (a) 0.8*mm* (b) 0.9*mm* (c) 1*mm*.

Figures 4 and 5 show the effect of buried electrode displacement on electron and ion density at 10ns. Increasing electrode distance decrease the number of electron and ion on the right side of the exposed electrode. The discharge starts at the right side of the exposed electrode and propagate to the wall over time. The maximum electron density in the edge of the exposed electrode is $1.4\times10^{18}\,(1/m^3)$.

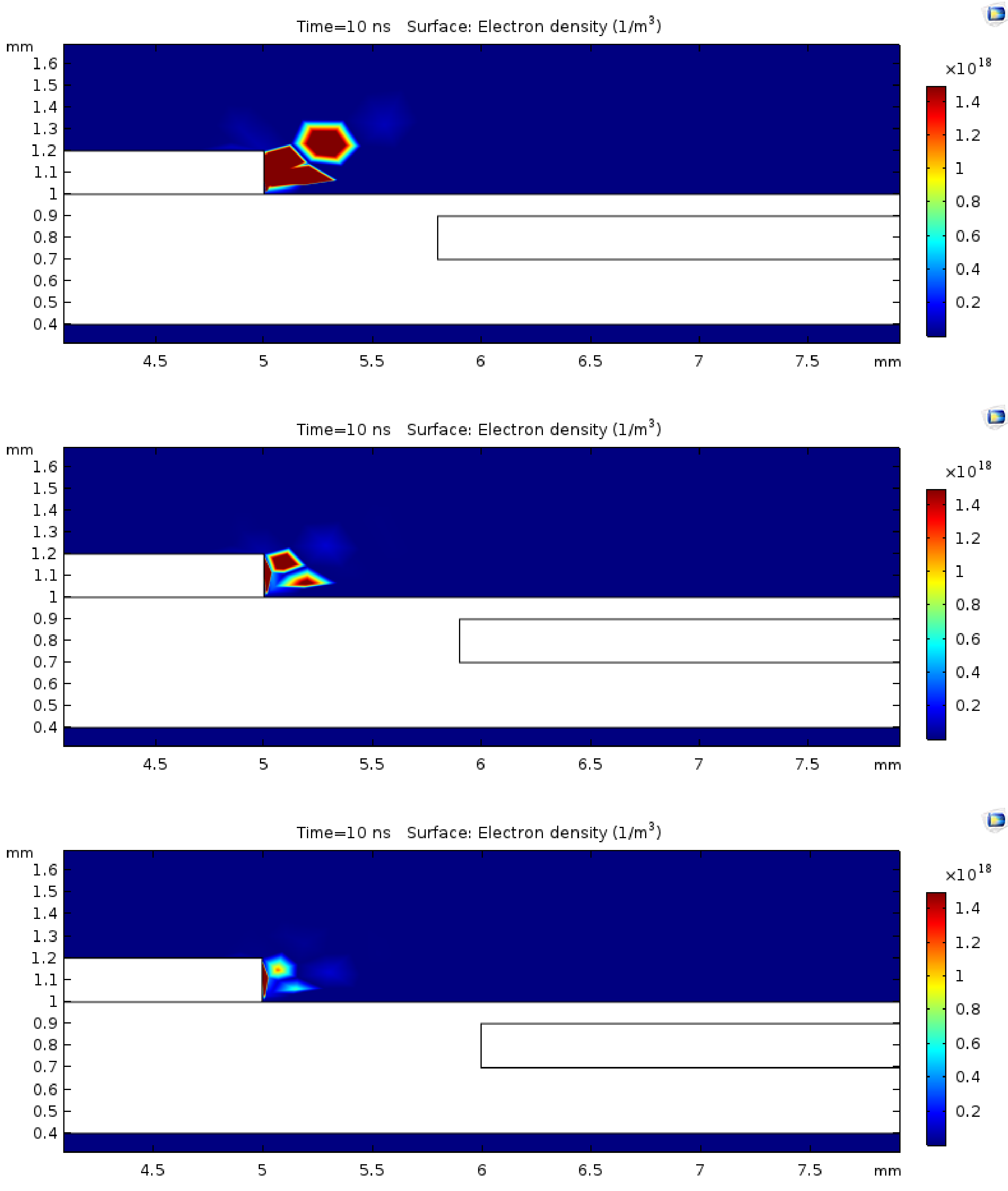

Fig4. The electron density at 1*kV* DC voltage over 10ns (a) 0.8*mm* (b) 0.9*mm* (c) 1*mm*.

The process of ion changes in this structure is more significant with the displacement of electrodes. The maximum electron density in the edge of exposed electrode is $2\times10^{19}\,(1/m^3)$.

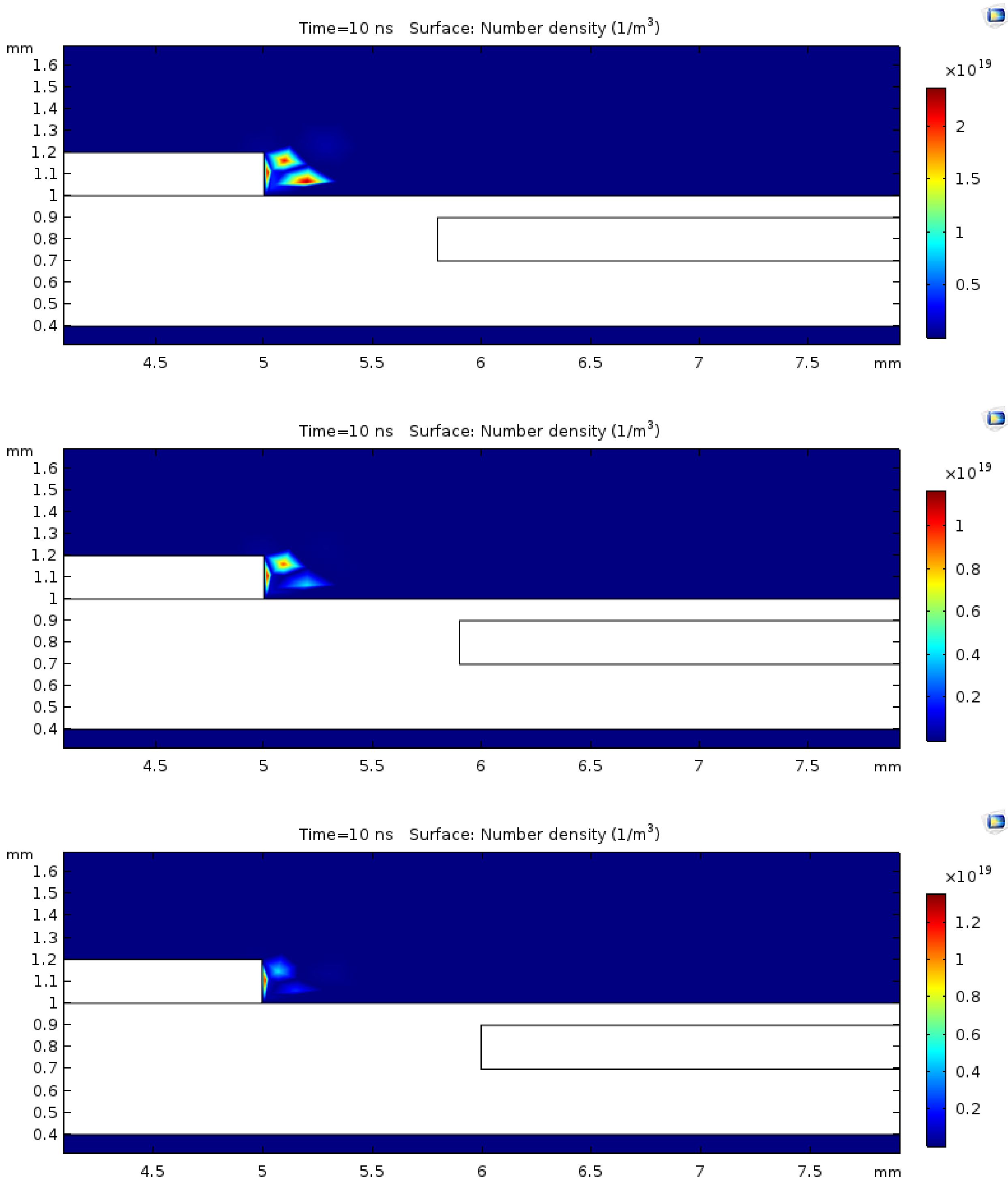

Fig5. The ion density at $1kV$ DC voltage over 10ns (a) 0.8*mm* (b) 0.9*mm* (c) 1*mm*.

For more investigations of the electron and ion density change due to the displacement of the buried electrode, electron and ion densities are shown on the upper side of the wall (fig 6). Fig 6 shows ion and electron density though a virtual line from points (5, 1.1) to (9 and 1.1) for different displacements of electrodes. Variation in density is more significant for distances of 0.8 *mm* to 0.9 *mm*.

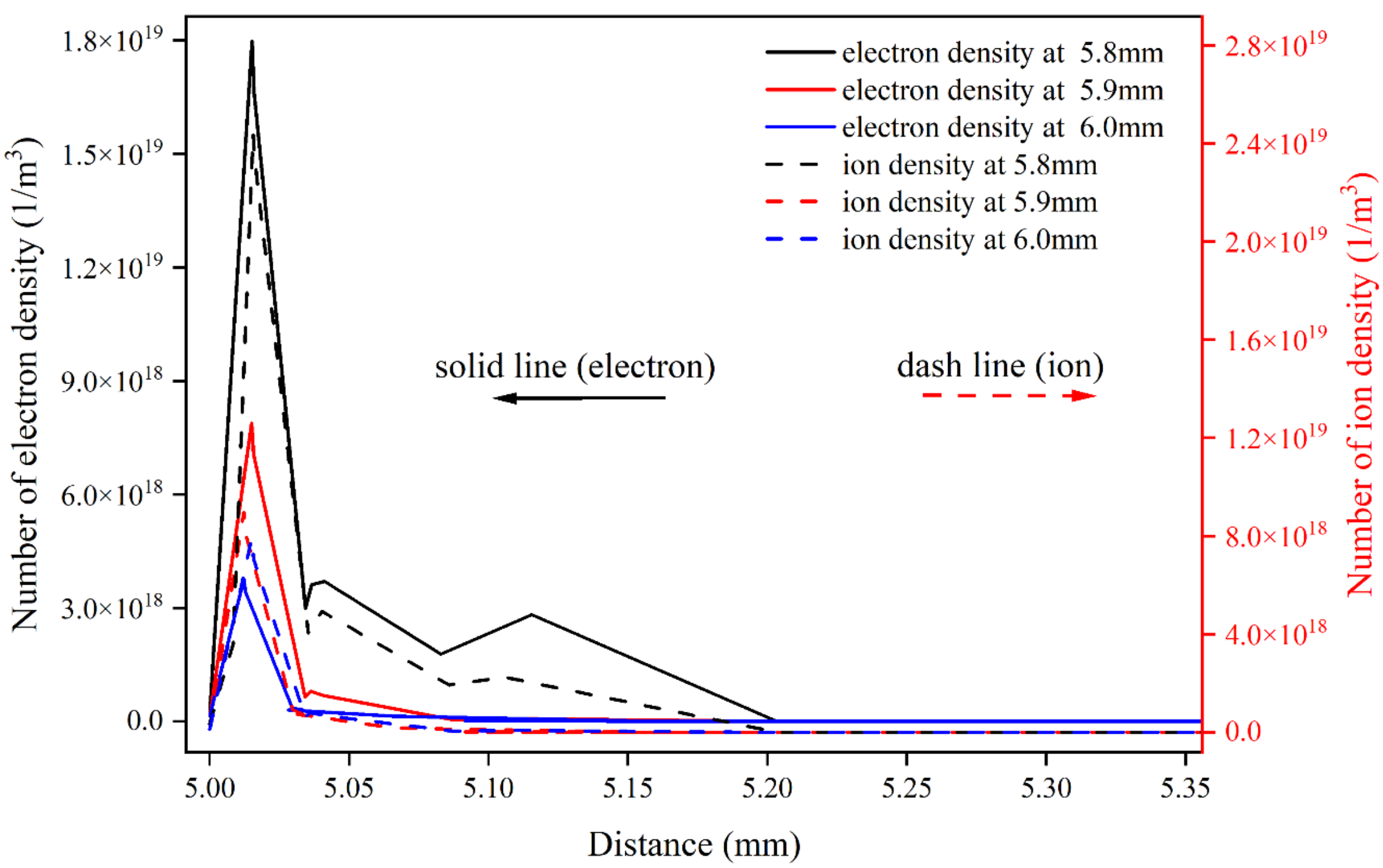

Fig6. Electron and ion density on the upper space of the wall.

Figure 7 shows the electron temperature distribution at 10 ns for different distances of the electrodes. The electron temperature near the exposed electrode is the maximum power and by moving away from it to the dielectric surface reaches less than 1 electron volt. The increase in the distance between the electrodes also reduces the electron temperature in the discharge zone.

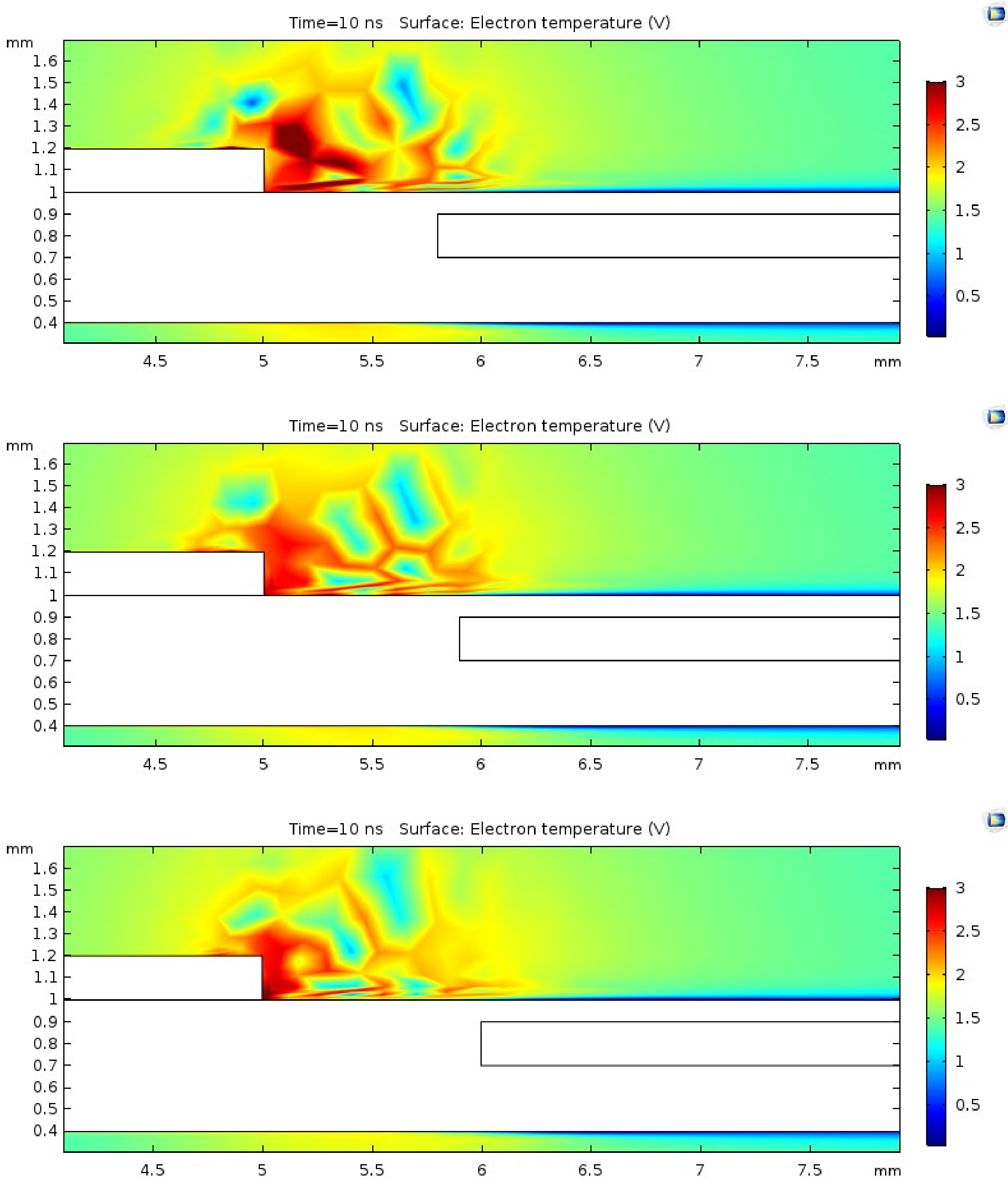

Fig7. Electron temperature distribution at discharge zone at 10ns (a) 0.8*mm* (b) 0.9*mm* (c) 1*mm.*

One of the effective parameters in creating body force is the electric field. Increasing the electric field apart from the direction is the main factor in increasing the body force. In Fig. 8, the density of the field lines and the field size in the space between the two electrodes and above the dielectric surface are shown.

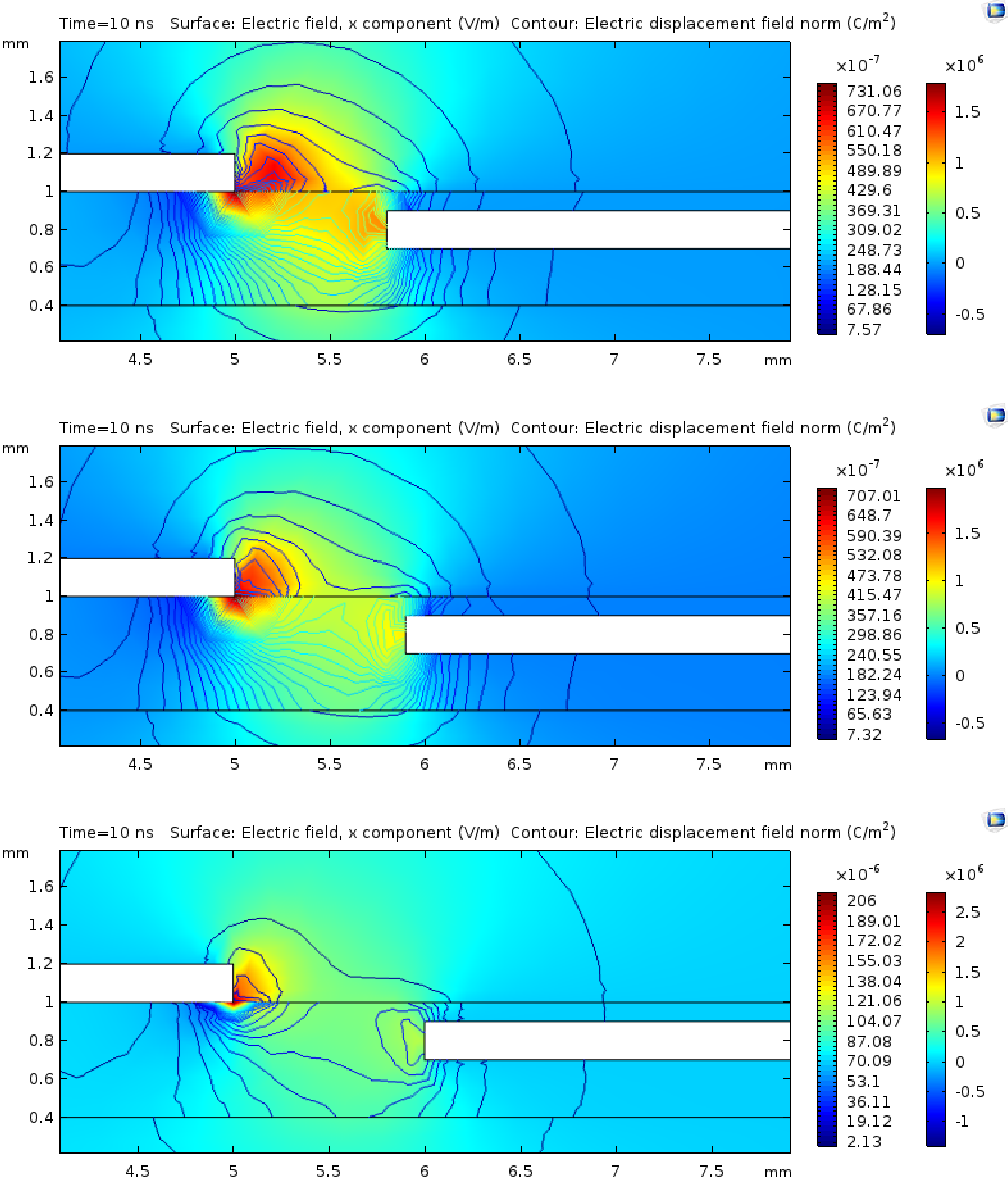

Fig 8. Density and magnitude of the electric field in plasma actuator (a) 0.8*mm* (b) 0.9*mm* (c) 1*mm.*

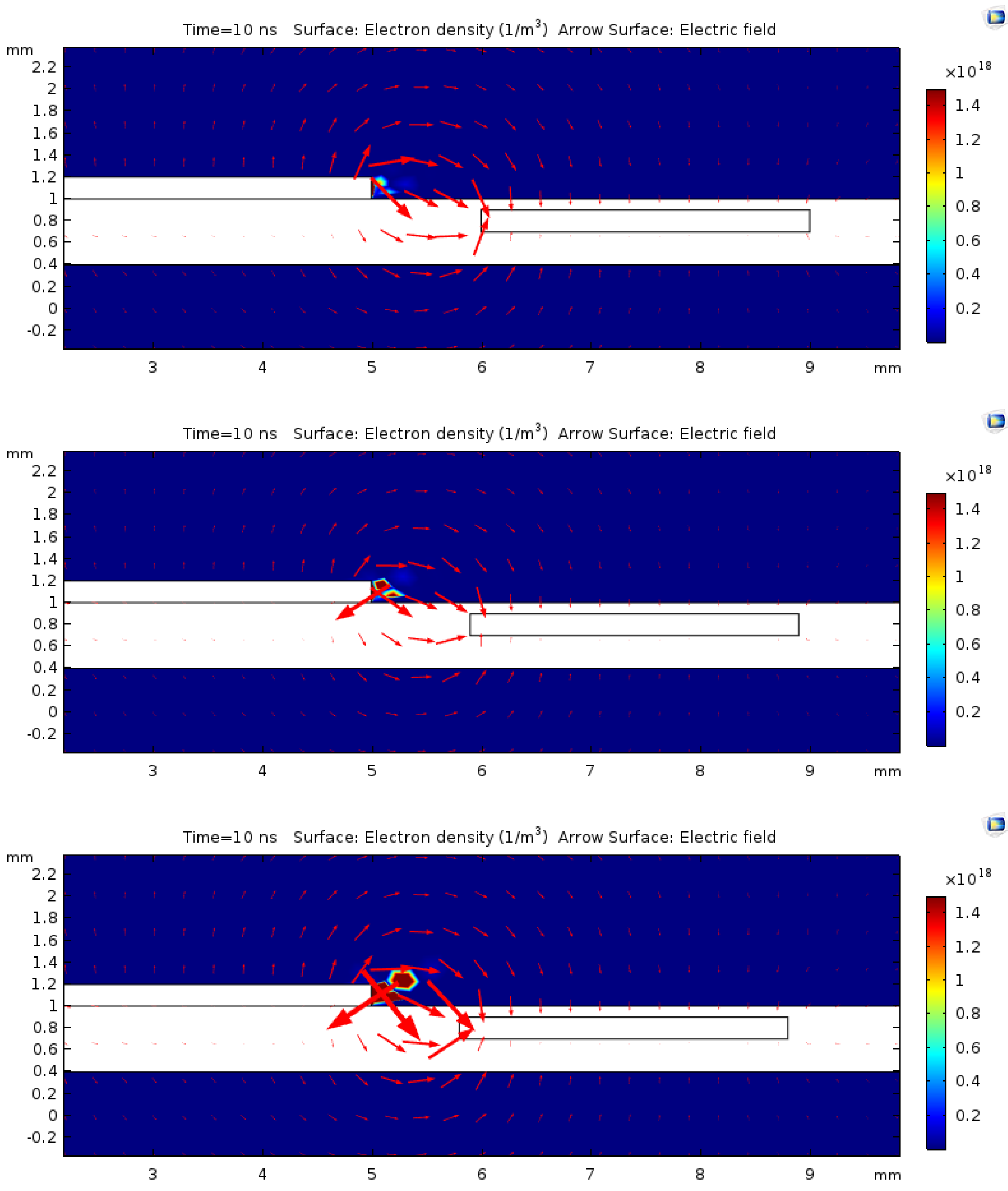

Fig 9. Electric field arrow made by electrode potential and charges (a) 0.8*mm* (b) 0.9*mm* (c) 1*mm*.

The dielectric material plays an important role in the rate of species production. Here are three different materials for the dielectric. Then the effect of these materials on electron and ion density has been investigated. The horizontal distance between the two electrodes is considered to be 1*mm* here, and the particle density at the moment of 10 nanoseconds is shown in Figure 10. Since the magnitude of the positive and negative charge difference has a direct effect on the magnitude of the body force, this value has been calculated for different dielectrics Figure 11. The highest charge difference is for mica dielectric and the lowest is for silica glass.

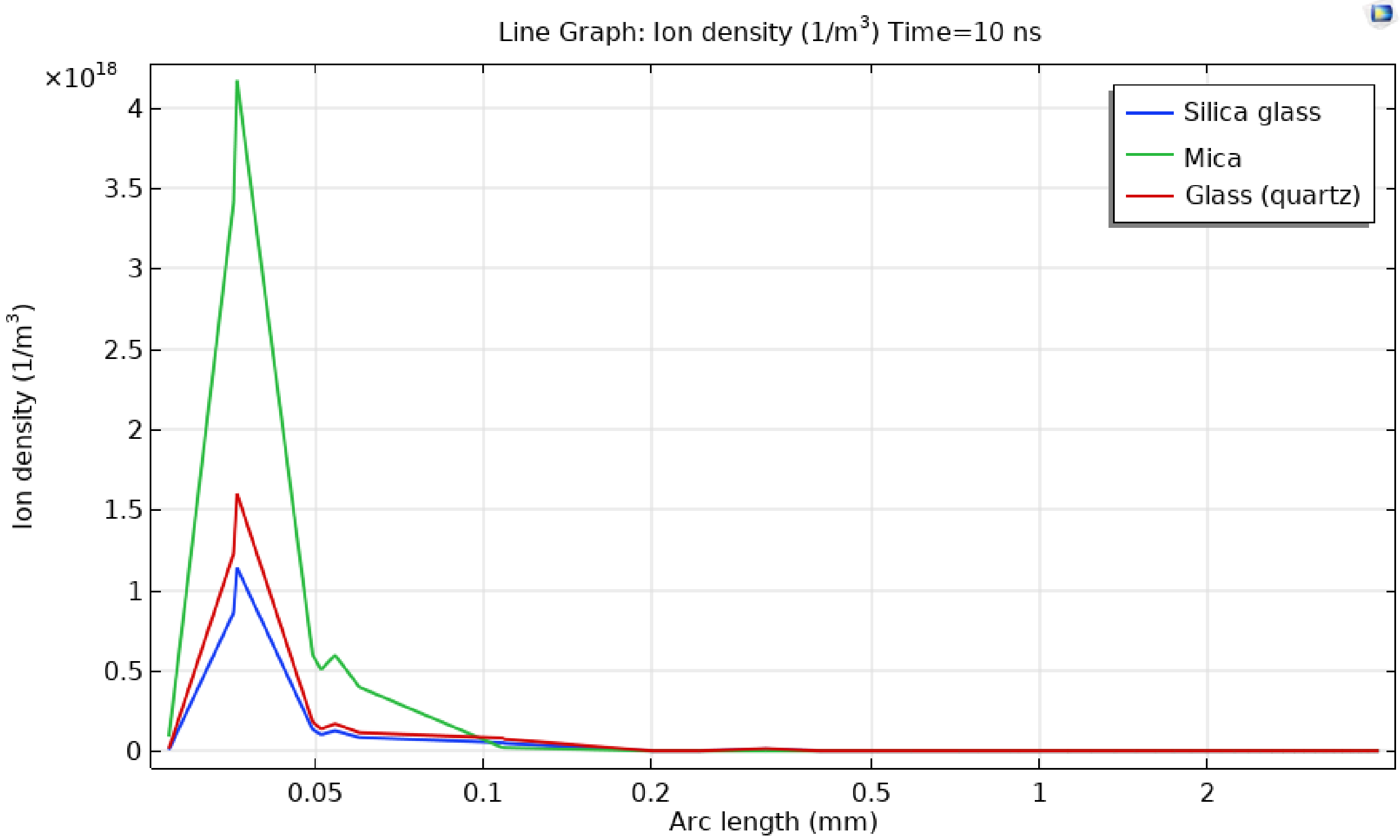

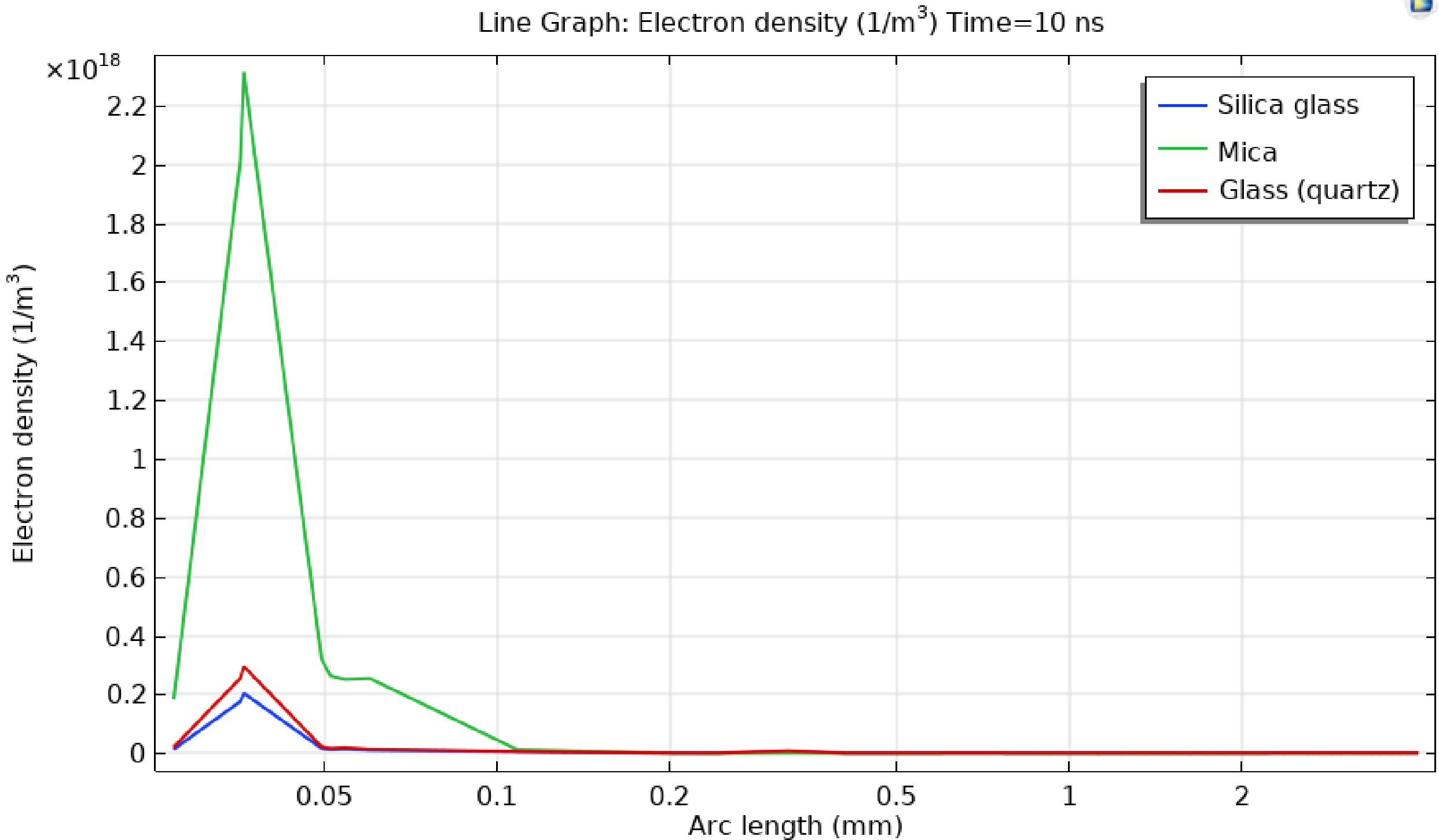

fig10. Ion (up) and electron (down) density in presence of different dielectric materials.

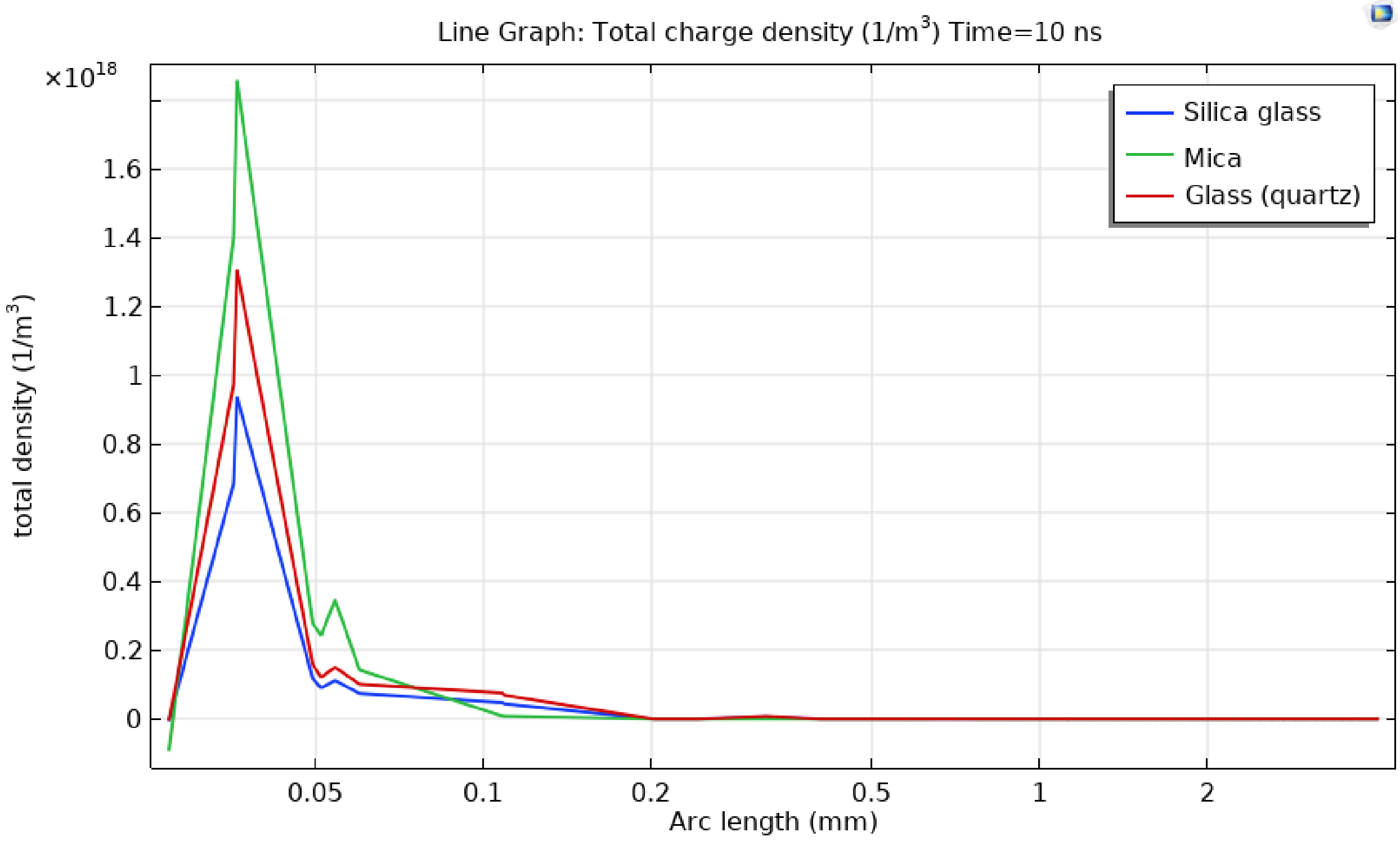

Fig 11. net charge production for different dielectric materials.

## Conclusion

In this study, the electric discharge of Argon gas in the two-dimensional structure of plasma actuators has been studied. The density of species at different intervals of the electrodes was studied. Also, in the discharge of argon gas, seven possible reactions and two surface reactions on the dielectric are considered. The magnitude of the field and net electric charge in the discharge zone are the main factor body force production. The results of the simulation of electrical discharge with Comsol show the significant effect of the buried electrode distance on the body force. Longitudinal displacement of 0.1*mm* (1.0-0.8*mm*) distance from the exposed electrode has a significant effect on plasma parameters, including ionic and electron and electric field. Also, as a dielectric, Mica can create the largest charge difference compared to Quartz and Glass.

## References.